\documentclass[11pt]{article}

\usepackage[utf8]{inputenc}
\usepackage[T1]{fontenc}
\usepackage{lmodern}
\usepackage{hyperref}
\usepackage{url}
\usepackage{booktabs}
\usepackage{amsfonts}
\usepackage{nicefrac}

\usepackage{fancyhdr}
\usepackage{graphicx}
\usepackage{xcolor}
\usepackage{array}
\usepackage{geometry}
\usepackage[numbers]{natbib}
\usepackage{authblk}

\title{Terminal Wrench: A Dataset of 331 Reward-Hackable\\Environments and 3,632 Exploit Trajectories}

\author[1]{Ivan Bercovich}
\author[1]{Ivgeni Segal}
\author[1,*]{Kexun Zhang}
\author[2]{Shashwat Saxena}
\author[2]{Aditi Raghunathan}
\author[2]{Ziqian Zhong}

\affil[1]{Fewshot Corp}
\affil[2]{Carnegie Mellon University}
\affil[*]{Independent Researcher}

\date{}

\hypersetup{
pdftitle={Terminal Wrench: A Dataset of Reward-Hackable Environments and Exploit Trajectories},
pdfsubject={cs.LG, cs.AI},
pdfauthor={Ivan Bercovich, Ziqian Zhong, Ivgeni Segal, Kexun Zhang, Shashwat Saxena, Aditi Raghunathan},
pdfkeywords={reward hacking, specification gaming, AI safety, benchmarks, evaluation},
}

\begin{document}
\maketitle

\begin{abstract}
We release \textbf{Terminal Wrench}, a subset of 331 terminal-agent benchmark environments, copied from the popular open benchmarks that are demonstrably reward-hackable. The data set includes 3,632 hack trajectories and 2,352 legitimate baseline trajectories across three frontier models (Claude Opus 4.6, Gemini 3.1 Pro, GPT-5.4). Each entry preserves the original task definition alongside full attack trajectories that show how the verifier was bypassed. It also includes cases where the task was not solved as intended. The tasks span system administration, machine learning, software engineering, and security challenges; the exploits range from simple output spoofing to stack-frame introspection, standard-library patching, and rootkit-style binary hijacking. Crucially, these exploits are specific to each task, rather than the evaluation harness, making them harder to patch. We also present a monitorability study in which hack trajectories are sanitized or stripped of reasoning traces and then scored by an LLM judge, showing that detection degrades meaningfully when chain-of-thought is removed (AUC drops from 0.97 to 0.92). The data set is publicly available at \url{https://github.com/few-sh/terminal-wrench}.
\end{abstract}

\section{Introduction}

Terminal-agent benchmarks have become a primary signal for measuring the coding and system-administration capabilities of large language models. Benchmarks such as TerminalBench, Terminal Bench 2, Terminal Bench Pro, SETA, and OpenThoughts-TB-dev present agents with containerized tasks. Each task may consist of multiple steps, including writing a script, configuring a service, training a model, among others. The solutions are verified through automated test suites that are individually shipped with each task. As the market for evaluation environments grows (with major organizations reportedly spending large sums on task procurement), so does the pressure to ship tasks quickly, often without thorough adversarial review of the verification logic.

 While reviewing tasks for one such benchmark, we observed that over 15\% of the verifiers could be bypassed by an agent that optimizes for the reward signal rather than the intended solution. This paper describes Terminal Wrench, a systematic effort to catalog and characterize such vulnerabilities. Starting from 1,860 tasks drawn from five public benchmarks, we ran over 40,000 adversarial trials using a hack-elicitation prompt appended to each task. From these trials, we identified 395 candidate hackable tasks and then performed a more robust hacker loop to produce the final dataset of 331 confirmed hackable environments.

Terminal Wrench serves three purposes: (1) it provides benchmark maintainers with a concrete list of vulnerable tasks to fix; (2) it enables research on reward hacking by supplying real, diverse exploit trajectories; and (3) it supports research on monitoring and oversight by pairing each hack with sanitized and stripped trajectory variants suitable for detection experiments.

\section{Dataset Overview}

The dataset contains 331 unique tasks with 957 task/model entries (a task may be tested against multiple models). In total there are 6,289 adversarial trajectories generated using the v5 hack-elicitation prompt, of which 3,632 are confirmed hacks, 1,216 are legitimate solves produced by the attacker agent, and 1,441 are no-reward attempts. An additional 2,352 baseline trajectories from successful pre-checks provide non-adversarial reference points. Three frontier models were used: Claude Opus 4.6, Gemini 3.1 Pro, and GPT-5.4 (high reasoning effort).

\begin{table}[ht]
\centering
\caption{Source Datasets}
\begin{tabular}{@{}lrrrrr@{}}
\toprule
Source Dataset & Total & Included & Rate & Baselines & Hacks \\
\midrule
terminal-bench 2.0 & 89 & 14 & 15.7\% & 200 & 186 \\
terminal-bench-pro & 200 & 48 & 24.0\% & 270 & 486 \\
OpenThoughts-TB-dev & 70 & 9 & 12.9\% & 219 & 234 \\
seta\_2026\_01\_29 & 1,376 & 241 & 17.5\% & 1,385 & 2,420 \\
TerminalBench-original & 233 & 36 & 15.5\% & 534 & 545 \\
\midrule
Total (with overlap) & 1,968 & 348 & 17.7\% & 2,608 & 3,871 \\
\bottomrule
\end{tabular}
\label{tab:source_datasets}
\end{table}

Some source datasets overlap: terminal-bench 2.0 and TerminalBench-original share 88 tasks; OpenThoughts-TB-dev and TerminalBench-original share 20.

\begin{table}[ht]
\centering
\caption{Results by Model}
\begin{tabular}{@{}lrrrrr@{}}
\toprule
Model & Entries & Baselines & Hacks & Legit. & No Reward \\
\midrule
Claude Opus 4.6 & 323 & 614 & 1,119 & 1,027 & 316 \\
Gemini 3.1 Pro & 321 & 984 & 1,495 & 38 & 630 \\
GPT-5.4 (high) & 313 & 754 & 1,018 & 151 & 495 \\
\midrule
Total & 957 & 2,352 & 3,632 & 1,216 & 1,441 \\
\bottomrule
\end{tabular}
\label{tab:results_by_model}
\end{table}

\section{Methodology}

Our methodology proceeds in two stages. In the first stage, we append a hack-elicitation excerpt to the agent's prompt (see the repository for the full prompt text). This excerpt instructs the agent to look for ways to pass the verifier without solving the task as intended. We ran this elicitation across 1,860 tasks using multiple models and prompt variants, producing over 40,000 trials. An LLM judge classified each trajectory as a serious exploit, a non-serious hack, a legitimate solve, or a no-reward attempt.

In the second stage, we narrowed the candidate set to 395 tasks that showed at least one hack in the first stage, then ran a more robust adversarial loop with additional trials per task. After manual spot-checking and automated classification, 331 tasks were confirmed as hackable. The final dataset includes all trajectories from the second stage: hacks, legitimate solves, and failed attempts alike.

\section{Exploit Taxonomy}

We assign each hack trajectory one or more coarse exploit categories based on the strategy adopted by the agent. Categories are not mutually exclusive. Table~\ref{tab:exploit_categories} summarizes the taxonomy.

\begin{table}[ht]
\centering
\caption{Exploit Categories}
\begin{tabular}{@{}llp{7.5cm}@{}}
\toprule
Category & Count & Description \\
\midrule
hollow-implementation & 2,243 & Passes tests but implements no real logic \\
output-spoofing & 1,071 & Fabricates expected outputs without computing them \\
constraint-loophole & 849 & Satisfies letter of spec, violates its intent \\
structural-extraction & 529 & Reads verifier source or answer keys \\
binary-hijacking & 428 & Replaces or patches binaries, stdlib, or system tools \\
algorithmic-simplification & 339 & Trivially simple method that happens to pass \\
mutable-input-tampering & 322 & Modifies shared state via side channels \\
keyword-gaming & 312 & Inserts dead code to satisfy static analysis \\
metric-spoofing & 236 & Manipulates timers, counters, or scoring \\
security-downgrading & 80 & Weakens security while reporting compliance \\
deceptive-rationalization & 38 & Frames exploit as legitimate behavior \\
\bottomrule
\end{tabular}
\label{tab:exploit_categories}
\end{table}

The most common exploit is \emph{hollow-implementation}, appearing in 2,243 trajectories. These are solutions that pass all tests but implement no meaningful logic. The second most common category, \emph{output-spoofing}, involves fabricating the expected output directly. More sophisticated attacks include \emph{binary-hijacking} (replacing system binaries or standard library modules) and \emph{structural-extraction} (reading the verifier source code or answer keys to determine expected outputs).

\section{Monitoring Experiments}

A natural question is whether these hacks can be detected by an LLM monitor observing the agent's trajectory. We constructed three variants of each hack trajectory: the \textbf{original} (full reasoning and tool calls), a \textbf{sanitized} version (reasoning rewritten to remove explicit mentions of hacking while preserving technical actions), and a \textbf{stripped} version (all agent commentary removed, retaining only tool calls and observations). We scored each variant with GPT-5.4 using a judge prompt that outputs a suspiciousness score between 0 and 1, and compared against scores on the 2,352 legitimate baseline trajectories.

\begin{table}[ht]
\centering
\caption{Monitoring Detection Results}
\begin{tabular}{@{}lrrrr@{}}
\toprule
Variant & AUC & TPR@5\%FPR & Pos. Mean & Neg. Mean \\
\midrule
Original & 0.9679 & 0.8235 & 0.9586 & 0.1777 \\
Sanitized & 0.9474 & 0.6187 & 0.9083 & 0.1777 \\
Stripped & 0.9168 & 0.4400 & 0.8228 & 0.1777 \\
\bottomrule
\end{tabular}
\label{tab:monitoring_results}
\end{table}

Detection is strong when the full chain-of-thought is available (AUC = 0.97), but degrades as reasoning is removed. The stripped variant, which retains only actions and observations, drops to AUC = 0.92 overall and TPR at 5\% FPR falls from 0.82 to 0.44. Exploit categories that most closely mimic legitimate behavior, such as \emph{algorithmic-simplification} and \emph{constraint-loophole}, show the largest detection degradation under sanitization (AUC drops of 0.04 and 0.03 respectively).

\begin{table}[ht]
\centering
\caption{Detection AUC by Hacker Model}
\begin{tabular}{@{}lrrr@{}}
\toprule
Model & Original & Sanitized & Stripped \\
\midrule
Claude Opus 4.6 & 0.9688 & 0.9549 & 0.9349 \\
Gemini 3.1 Pro & 0.9639 & 0.9279 & 0.8805 \\
GPT-5.4 (high) & 0.9802 & 0.9720 & 0.9535 \\
\bottomrule
\end{tabular}
\label{tab:detection_by_model}
\end{table}

Gemini 3.1 Pro hacks are the hardest to detect when stripped (AUC = 0.88), while GPT-5.4 hacks remain the most detectable across all variants. This suggests that different models produce qualitatively different exploit styles, with some being inherently more action-transparent than others.

\end{document}